\documentclass[11pt,a4paper, final, twoside]{article}
\usepackage{amsmath}
\usepackage{fancyhdr}
\usepackage{amsthm}
\usepackage{amsfonts}
\usepackage{amssymb}
\usepackage{amscd}
\usepackage{latexsym}
\usepackage{amsthm}
\usepackage{graphicx}
\usepackage{graphics}
\usepackage{natbib}
\usepackage{afterpage}
\usepackage[colorlinks=true, urlcolor=blue,  linkcolor=black, citecolor=black]{hyperref}
\usepackage{color}

\newtheorem{thm}{Theorem}[section]
\newtheorem{algorithm}[thm]{Algorithm}

\newtheorem{example}[thm]{Example}

\theoremstyle{proposition}

\theoremstyle{definition}
\newtheorem{defn}{Definition}[section]
\theoremstyle{remark}

\numberwithin{equation}{section}
\begin{document}
\hyphenpenalty=100000

\begin{flushright}
{\Large \textbf{Bitwise Operations in Relation to Obtaining Latin Squares}}\\[5mm]
{\large \textbf{Krasimir Yordzhev$^\mathrm{1^*}$\footnote{\emph{*Corresponding author: E-mail: yordzhev@swu.bg;}}  }}\\[3mm]
$^\mathrm{1}${\footnotesize \it Faculty of Mathematics and Natural Sciences, South-West University,
2700 Blagoevgrad, Bulgaria}\\[3mm]

\footnotesize
\end{flushright}
{\rule{\linewidth}{1pt}}\\[4mm]
\footnotesize
{\Large \textbf{Abstract}}\\[4mm]
\fbox{%
\begin{minipage}{5.4in}{\footnotesize The main thrust of the article is to provide interesting example, useful for students of using bitwise operations in the programming languages C ++ and Java.  As an example, we describe an algorithm for obtaining a Latin square of arbitrary order. We will outline some techniques for the use of bitwise operations.\\
} \end{minipage}}\\[3mm]
\footnotesize{\it{Keywords:} Latin square; integer representation of sets; exponential Latin square; bitwise operations.}\\[3mm]
\footnotesize{\bf{2010 Mathematics Subject Classification:}} 68N15.

\section{Introduction}

Let $n$ be a positive integer. Throughout $[n]$ denotes the set
$$[n] =\left\{ 1,2,\ldots ,n\right\} .$$
Let $A\subseteq [n]$. We denote by $\mu_i (A)$ the functions
\begin{equation}\label{mu}
\mu_i (A) =\left\{
\begin{array}{ccc}
  1 & \textrm{if} & i\in A \\
  0 & \textrm{if} & i\notin A
\end{array}
\right.
,\ i=1,2,\ldots .
\end{equation}
Then the set $A$ can be represented uniquely by the integer
\begin{equation}\label{nu}
\nu (A) =\sum_{i=1}^n \mu_i (A) 2^{i-1} , \quad 0\le \nu (A)\le 2^n -1 ,
\end{equation}
where $\mu_i (A)$, $i=1,2,\ldots ,n$ is given by formula (\ref{mu}). In other words, we will represent each subset of $[n]$ uniquely with the help of an integer from the interval $[0,2^n -1]$ (\emph{integer representation of sets}).\\[3mm]
It is readily seen that
\begin{equation}\label{U}
\nu ([n])=2^{n} -1.
\end{equation}
Evidently if $A=\{ a\}$, i.e. $|A|=1$, then
\begin{equation}\label{nua}
\nu (\{ a\})=2^{a-1}.
\end{equation}
The empty set $\emptyset$ is represented by
\begin{equation}\label{emptys}
\nu (\emptyset ) =0.
\end{equation}
\begin{defn}
A \emph{Latin square} of order $n$ is an $n \times n$ matrix where each row and each column is a permutation of elements of $[n]$.
\end{defn}
\noindent Thus far it is known the number of all Latin squares of order $n$, where $n\le 11$ [\cite{McKay}, \cite{A002860}].\\[3mm]
Latin squares and hypercubes have their applications in coding theory, error
correcting codes, information security, decision making, statistics, cryptography,
conflict-free access to parallel memory systems, experiment planning, tournament
design, enumeration and study of H-functions, etc [\cite{Laywine}, \cite{Kovachev}].

\begin{defn}
A matrix $M_{n\times n}=\left( \alpha_{ij} \right)_{n\times n}$ is called an \emph{exponential Latin square} of order $n$ if the following conditions hold:
\begin{enumerate}
  \item For every $i,j\in [n]$ there exists $k\in [n]$ such that $\alpha_{i,j} =2^k$;
  \item The matrix $M_{n\times n}' = \left( \log_2 \alpha_{ij} \right)_{n\times n} $ is a Latin square of order $n$.
\end{enumerate}
\end{defn}
\noindent  If $n$ is a positive integer, then it is readily seen that the set of all $n\times n$ Latin squares and the set of all $n\times n$ exponential Latin squares are  isomorphic.\\[3mm]
The use of bitwise operations is a powerful method used in C/C++ and Java programming languages. Unfortunately, in the widespread books on this topic there is incomplete or no description for the work of bitwise operations. The aim of this article is to correct this lapse to a certain extent and present a meaningful example of a programming task, where the use of bitwise operations is appropriate in order to facilitate the work and to increase the effectiveness of the respective algorithm.\\[3mm]
Other interesting examples of effective use of bitwise operations can be found in publications [\cite{YoKo}], [\cite{sort}] and [\cite{bin}].\\[3mm]
In this paper we will show that it is easy to create an algorithm for generating random exponential Latin squares of order n using bitwise operations. The presentation of the subsets of the set $[n]$ using the formula (\ref{nu}) and the convenience in this case to work with bitwise operations are the basis of the algorithm described by us. Some other algorithms for obtaining random Sudoku matrices and their valuation are described in detail in [\cite{DeSalvo}], [\cite{Fontana}] and  [\cite{yordzhev_random}].
\section{Bitwise Operations}\label{bitwiseop}
The use of bitwise operations is a well-working method, used in programming languages  C/C++ and Java. Let us recall the basic definitions related to bitwise operations. More details  could be seen, for example, in [\cite{2}], [\cite{5}], [\cite{Magda}] for the C/C++ programming languages and in [\cite{4}], [\cite{7}] for the Java programming language.\\[3mm]
Bitwise operations can be applied for integer data type only. We assume as usual that bit numbering in variables starts from right to left, and that the number of the rightmost one is 0.\\[3mm]
Let  \verb"x,y"  and \verb"z" be integer variables or constants of one type, for which $w$  bits are needed. Let \verb"x"  and  \verb"y" be initialized (if they are variables) and let the  assignment \verb"z = x&y;" (\emph{bitwise AND}), or \verb"z = x|y;" (\emph{bitwise inclusive OR}), or \verb"z = x^y;" (\emph{bitwise exclusive OR}), or \verb"z = ~x;" (\emph{bitwise NOT}) be made. For each $i=0,1,2,\ldots ,w-1$,  the new contents of the $i$-th  bit in \verb"z"  will be as it is presented in the Table {1}.\\[3mm]
\begin{table} 
\footnotesize
\begin{center}
{\bf{ Table 1. Bitwise operations}\label{bitwise}}\\[3mm]
\begin{tabular}{|c|c|c|c|c|c|}
  \hline
 $i$-th bit of &  $i$-th bit of &  $i$-th bit of  & $i$-th bit of  &  $i$-th bit of  &  $i$-th bit of  \\
 \verb"x" &  \verb"y" & \verb"z = x&y;" & \verb"z = x|y;" & \verb"z = x^y;" & \verb"z = ~x;"\\
\hline\hline
  0 & 0 & 0 & 0 & 0 & 1\\ \hline
  0 & 1 & 0 & 1 & 1 & 1\\ \hline
  1 & 0 & 0 & 1 & 1 & 0\\ \hline
  1 & 1 & 1 & 1 & 0 & 0\\ \hline
\end{tabular}
\end{center}
\end{table}
\noindent
In the case that  \verb"k"  is a nonnegative integer, the statement \verb"z = x<<k"  (\emph{bitwise shift left}) will write $(i+k)$ in the bit of \verb"z" the value of the $k$th bit of \verb"x", where $i=0,1,\ldots ,w-k-1$, and the rightmost $k$  bits of \verb"x" will be filled by zeroes. This operation is equivalent to a multiplication of  \verb"x"  by  $2^k$.\\[3mm]
The statement \verb"z=x>>k" (\emph{bitwise shift right}) works in a similar way. But we must be careful if we use the programming language C or C++, as in various programming environments this operation has different interpretations – sometimes the leftmost $k$ bits of \verb"z"  are compulsorily filled by 0 (logical displacement), and elsewhere the leftmost $k$  bits of \verb"z"  are filled with the value from the leftmost (sign) bit; i.e. if the number is negative, then the filling will be with 1 (arithmetic displacement). Therefore it's recommended to use \verb"unsigned" type of variables (if the opposite is not necessary) while working with bitwise operations. In the Java programming language, this problem is solved by introducing two different operators: \verb"z=x>>k" and \verb"z=x>>>k".

\begin{example}
To compute the value of the $i$-th  bit of an integer variable \verb"x" we can use the function:
\begin{verbatim}
int BitValue(int x, unsigned int i) {
	return ( (x & 1<<i) == 0 ) ? 0 : 1;
}
\end{verbatim}
\end{example}

\begin{example}
The next function displays an integer in binary notation. We don't consider and we don't display the sign of the integer. For this reason we work with $|n|$.
\begin{verbatim}
void DecToBin(int n)
{
     n = abs(n);
     int b;
     int d = sizeof(int)*8 - 1;
     while ( d>0 && (n & 1<<(d-1) ) == 0 ) d--;
     while (d>=0)
           {
           b= 1<<(d-1) & n ? 1 : 0;
           cout<<b;
           d--;
           }
}
\end{verbatim}
\end{example}

\begin{example}
The following function calculates the number of 1's in the binary representation of an integer n.
Again we ignore the sign of the number.
\begin{verbatim}
int NumbOf_1(int n)
{
     n = abs(n);
     int temp=0;
     int d = sizeof(int)*8 - 1;
     for (int i=0; i<d; i++)
         if (n & 1<<i) temp++;
     return temp;
}
\end{verbatim}
\end{example}

\begin{example}
Bitwise operations and sets (See also [\cite{KoYo}]). We will use expressions that represent operations with sets in the next section for a description of the algorithm to obtain a random exponential Latin square.
\begin{enumerate}
  \item According to formula (\ref{U}), the set $U=[n]=\{ 1,2,\ldots \}$ is represented by the expression $$\verb"U = (1<<n) - 1;"$$
  \item According to formula (\ref{nua}), a singleton $A=\{ a\} \subset [n]$ is represented by the expression $$\verb"A = 1<<(a-1);"$$
  \item Let $A$ and $B$ be two integers, which represent two subsets of $[n]$. Then the integer $C$ that represents their union is represented by the expression $$\verb"C = A | B;"$$
  \item Let $A\subseteq B$. Then, to remove all elements of $A$ from $B$, we can do it with the help of the expression $$\verb"B ^ A".$$
\end{enumerate}
\end{example}

\section{Description and Implementation of the Algorithm}

We will describe an algorithm created by us, giving the source code written in the C++ programming language with detailed comments.
\begin{algorithm}\label{alg1}

\begin{verbatim}

#include <iostream>
#include <cstdlib>
#include <ctime>
#define N 12

using namespace std;

int U = (1<<N) - 1;
    /* U represents the universal set [n] = {1,2, ... ,N}
        according to (1.3) */
int L[N][N];
    /* L- exponential Latin square */

int choice(int k) {	
    /* In this case, "choice" means random choice of a 1
       from the binary notation of positive integer k, 0<k<=U */
    if (k<=0 || k>U) {
        cout<<"The choice is not possible \n";
        return 0;
    }
    int id =0;
        /* id - the number of 1 in the binary notation of k */
    for (int i=0; i<N; i++)
        if (k & 1<<i) id++;
    srand(time(0));
    int r = rand()%id+1;
        /* r is a random integer, 1<=r<=id. The function will
            chooses the r-th 1 from the binary notation of k.*/
    int s=0, t=1;
    while (1) {
        if (k&t) s++;
            /* s-th bit of the integer k is equal to 1. */
        if (s==r) return t;
            /* The function returns the integer t=2^{r-1},
                which represents the singleton {r}. */
        t=t<<1;
            /* t=t*2 and check the next bit */
    }
}

void use() {	
        /* In this case "use" means "printing" */
    for (int i=0; i<N; i++){
        for (int j=0; j<N; j++)
            cout<<L[i][j]<<" ";
        cout<<endl;
    }
}

int main(int argc, char** argv) {
    int row,col;
    int A;
        /* A represents a subset of [n] according to (1.2) */
    for (row=0; row<N; row++) {
        col =0;
        while (col<N) {
            A=0; /* empty set */
            for (int i=0; i<row; i++) A = A | L[i][col];
            for (int j=0; j<col; j++) A = A | L[row][j];
            A = U ^ A;	
                /* The algorithm will select an element of this set
                    and insert it into the next position. */
            if (A!=0) {
                L[row][col] = choice(A);
                col++;
            }
            else col = 0;
        }
    }	
    use();
return 0;
}
\end{verbatim}
\end{algorithm}

\section{Conclusion}
With the help of algorithm \ref{alg1}, we received a lot of random exponential Latin squares, for example the next of order 12:
$$
\begin{array}{llllllllllll}
32 & 1 & 16 &	8 &	512 &	256 &	2048 &	128 &	2 &	1024 &	4 &	64 \\
4 &	256 &	512 &	16 &	8 &	64 &	1024 &	2 &	32	& 1	& 2048 &	128 \\
8 &	32 &	1 &	2 &	64 &	4 &	16 &	256 &	128 &	512 &	1024 &	2048 \\
16 &	4 &	8 &	32 &	2048	& 1024 &	512 &	64 &	256 &	128 &	2	& 1 \\
256 &	128 &	32 &	64 &	4 &	8 &	1 &	16 &	2048 &	2 &	512 &	1024 \\
512 &	2048 &	1024 &	256 &	1	& 128 &	2 &	32 &	64 &	16 &	8 &	4 \\
1024 &	16	& 256 &	128 &	32 &	2048 &	8	& 4	& 512 &	64 &	1 &	2 \\
2 &	64 &	128 &	4 &	1024 &	16 &	256 &	8 &	1 &	2048 &	32 &	512 \\
1 &	2 &	4 &	512 &	16 &	32 &	128 &	2048 &	1024 &	256 &	64 &	8 \\
64 &	8 &	2048 &	1024 &	2	& 512 &	32 &	1 &	16 &	4 &	128 &	256 \\
2048 &	1024 &	64 &	1 &	128 &	2 &	4 &	512 &	8 &	32 &	256 &	16 \\
128 &	512 &	2 &	2048 &	256 &	1	& 64	& 1024 &	4 &	8 &	16 &	32
\end{array}
$$
\noindent{\Large\bf Competing Interests}\\\\\
Author has declared that no competing interests exist.

\end{document}